\documentclass[12pt]{article}

\usepackage{graphicx}
\usepackage{epsf}
\usepackage{amsmath2000}
\allowdisplaybreaks

\begin{document}

\baselineskip=17.5pt plus 0.2pt minus 0.1pt

\makeatletter
\@addtoreset{equation}{section}
\def\CR{\nonumber \\}
\def\pt{\partial}
\def\be{\begin{equation}}
\def\ee{\end{equation}}
\def\bea{\begin{eqnarray}}
\def\eea{\end{eqnarray}}
\def\eq#1{(\ref{#1})}
\def\la{\langle}
\def\ra{\rangle}
\def\hyp{\hbox{-}}

\begin{titlepage}
\title{
\hfill\parbox{4cm}
{ \normalsize YITP-02-60 \\{\tt hep-th/0209220}}\\
\vspace{1cm}
Dualities of the entropy bound}
\author{
Naoki {\sc Sasakura}\thanks{\tt sasakura@yukawa.kyoto-u.ac.jp}
\\[15pt]
{\it Yukawa Institute for Theoretical Physics, Kyoto University,}\\
{\it Kyoto 606-8502, Japan}}
\date{\normalsize September, 2002}
\maketitle
\thispagestyle{empty}

\begin{abstract}
\normalsize
I study the hypothetical thermodynamic system which saturates the so-called 
Hubble entropy bound and show that it is invariant 
under the S- and T-dualities of string theory as well as the interchanges 
of the eleventh dimension of M-theory.
I also discuss how unique the entropy bound is under the dualities and 
some related issues.
\end{abstract}
\end{titlepage}

It was argued by Atick and Witten \cite{Atick:1988si} 
that the high temperature behavior
of the string theory free energy with volume $V$ and temperature $T$ 
is given by \footnote{In this paper,
for notational simplicity, I omit the $\alpha'$ parameter. I do not also 
care about inessential numerical factors. On the other hand, I take 
care much about dilaton dependence.} 
\be
\label{stringfree}
F_{string}\sim -VT^2.
\ee
They compared this behavior with the quantum field theory behavior 
in $d$ spatial dimensions,
\be
F_{qft}\sim -VT^{d+1},
\ee
and gave the interpretations that the fundamental degrees of freedom 
in string theory 
are much less than those of quantum field theory in the same dimensions, 
and that the theory underlying string theory should be like 
a $(1+1)$-dimensional quantum field theory.
This dramatic interpretation is fascinating and valuable to pursue
further, but the perturbative method of string theory has been limited by the 
existence of the Hagedorn transition and, despite various efforts, 
the validity of this conjecture has been remaining unclear. Nevertheless,
there is also another interesting observation to understand the behavior 
\eq{stringfree} \cite{Atick:1988si,Polchinski:rq}. 
Assuming the T-duality along the compact 
temperature direction\footnote{Because of the complications of the boundary 
conditions of fermions in the temperature direction, the T-duality of 
temperature does not hold in the naive sense as above in superstring theory. 
Also it was argued that the duality symmetry should be spontaneously broken
in the heterotic string theory \cite{Atick:1988si}. Hence 
the above discussions of T-duality should be regarded as a formal 
discussion to intuitively understand how the behavior \eq{stringfree} 
is characteristic of string theory.} and 
equating the partition functions of the both sides of the T-duality, 
the following equation holds;
\be
\label{equalpart}
\frac{1}{T}F_{string}(T)=\frac{T}{T_H^2}F_{string}(T_H^2/T),
\ee   
where $T_H$ is the Hagedorn temperature. The high temperature behavior 
\eq{stringfree} is obtained from \eq{equalpart}
by assuming the non-vanishing of $F_{string}(0)$.

A hypothetical thermodynamic system with a similar temperature behavior of 
free energy can be obtained from the entropy bound which has recently appeared.
A universal entropy bound dates back from the proposal of Bekenstein 
\cite{Bekenstein:jp} that the total entropy of an isolated system 
with energy $E$ and scale $R$ should be bounded by $ER$. This entropy bound is 
a quantity conserved under the adiabatic change of thermal radiation matters, 
and therefore
has a good reason to believe that it be applicable generally to quantum 
field theory.
Moreover, combined with the Schwarzschild bound $GE<R^{d-2}$, 
the entropy bound becomes $R^{d-1}/G$, 
which agrees with the holography hypothesis 
\cite{'tHooft:gx,Susskind:1994vu}\footnote{On more precise discussions about
the relations between 
the entropy bounds and the holography hypothesis, see \cite{Flanagan:1999jp}.}.
On the other hand, the entropy bound which appeared recently has, among
various formulations, essentially the form\footnote{The Hubble entropy bound 
is given by $S_H=HV/G$, 
where $H$ is the Hubble parameter \cite{Veneziano:1999ty}. 
Using the Einstein equation $H^2\sim G E/V$, the above
expression \eq{boundev} is derived.} \cite{Sasakura:1999xp}
\be
\label{boundev}
S=\sqrt{\frac{EV}{G}},
\ee
where $G$ is the gravitational coupling constant.
In string theory, this formula is written as 
\be
\label{entropybound}
S=e^{-\phi}\sqrt{EV},
\ee
where $\phi$ is the dilaton with the string coupling constant defined by
$g=e^\phi$.
Considering a hypothetical thermodynamic system 
which saturates the entropy bound 
\eq{entropybound} and using the first law of thermodynamics, its free energy 
becomes
\be
\label{freeenergy}
F=-e^{-2\phi}VT^2,
\ee
where I have neglected an inessential numerical factor.

The difference between \eq{freeenergy} and the string theory free energy 
\eq{stringfree} is the dilaton dependence. This difference is essential.
In perturbative string theory, to obtain the temperature dependent
part of the free energy, 
it is necessary for a string world sheet to wind around the 
compact temperature direction, and there is no genus-zero 
contribution like \eq{freeenergy}. Moreover
the motivations for the entropy 
bound \eq{boundev} come from cosmological considerations 
\cite{Fischler:1998st,Bousso:1999xy,Veneziano:1999ty}, a space-time
uncertainty relation based on general relativity and quantum mechanics
\cite{Sasakura:1999xp}, 
and a generalization of (curved space)/CFT correspondence 
\cite{Verlinde:2000wg}. The hypothetical matter satisfying \eq{freeenergy}
is the stiffest matter and is the main object in the cosmology of 
\cite{Banks:2001px}.
Thus the entropy bound \eq{entropybound} originated from these rather
macroscpic considerations, and there seem to be no reasons to think about the 
possibility of deriving 
\eq{freeenergy} from any microscopic computations of string theory. 
Nevertheless, in this paper, I will point out that
the formula \eq{freeenergy} satisfies the
duality symmetries of string theory and a requirement from M-theory. 
Assuming the duality invariance and that the free energy be 
proportional to the volume, the formula will be shown to be unique. 
These properties are quite amusing, suggesting that \eq{freeenergy} might
have a microscopic origin in string theory. In fact, in the original
paper by Atick and Witten \cite{Atick:1988si}, 
the possibility of genus-zero contributions 
above the Hagedorn transition was pointed out.
I will discuss this possibility further at the end of this paper.

To show the duality invariance of the free energy \eq{freeenergy}, let me 
recapitulate the T- and S- duality transformations of string theory 
\cite{T-duality,Polchinski}.
I take the space to be a periodic box (torus) of nine dimensions, and  
the string metric is assumed to be of the form
\be
ds_{string}^2=-e^{2\psi}dt^2 + \sum_{i=1}^9 e^{2\lambda_i} dx_i^2,
\ee
where the metric depends only on time.
I also assume that the dilaton field depends only on time. 
Then the string theory gravity-dilaton effective action in the 
lowest order approximation is given by 
\bea
S_{string}&=&-\int dt d^9x \sqrt{-g}e^{-2\phi}(R+4(D\phi)^2)\cr
&=& -V_9\int dt\ e^{-\psi-\varphi}
(\sum_{i=1}^9 \dot \lambda_i^2-\dot\varphi^2),
\eea
where $V_9$ is the spatial volume in the unit of $x$, and 
\be
\varphi=2 \phi-\sum_{i=1}^9\lambda_i.
\ee
This string theory effective action is invariant under the following T-duality 
transformations for a direction $i$,
\bea
\label{tduality}
\lambda_i&\rightarrow& -\lambda_i,\cr
\phi&\rightarrow& \phi-\lambda_i,\cr
\psi&\rightarrow& \psi,
\eea
and the S-duality transformation 
\bea
\label{sduality}
\phi&\rightarrow& -\phi, \cr
\lambda_i&\rightarrow&\lambda_i-\frac{\phi}{2}\ \ ({\rm for\ all}\ i), \cr
\psi&\rightarrow& \psi-\frac{\phi}{2}.
\eea 

Following the method of \cite{Tseytlin:1991xk}, 
the effective action for the thermodynamic system associated to 
\eq{freeenergy} is given by 
\bea
\label{hypoaction}
S_{hyp}&=&\int dt \sqrt{-g_{00}}F(\phi,\lambda_i,\beta \sqrt{-g_{00}}), \cr
&=&- V_9 \int dt\ e^{-\psi-\varphi}\ T^2.
\eea
where $\beta$ is the inverse of the temperature. The $g_{00}$ dependences are 
necessary for the reparameterization invariance of the time direction. The 
temperature $T$ here is measured in the unit of the inverse of time $t$, 
and is therefore 
invariant under the duality transformations, which are defined as 
transformations on the background fields.
The combination $-\psi-\varphi$ is obviously invariant under the dualities
\eq{tduality} and \eq{sduality}, and so is the action \eq{hypoaction}.

Now let me start from the assumption that the free energy be proportional
to the volume. Then the T-duality transformation \eq{tduality} requires that
the dilaton field must be combined with the volume in the form 
$\varphi=2\phi-\sum_{i=1}^9 \lambda_i$. At this point I am not demanding that
another dilaton dependence is impossible in string theory. 
If I consider another form
of the dilaton dependence, I will need other terms with
non-local $1/R_i=e^{-\lambda_i}$ behaviors as well as the local 
$R_i=e^{\lambda_i}$ ones. 
Because of the winding modes of string, the non-local behaviors 
certainly appear when the compact directions are in the order of the 
string scale. Nevertheless the free energy \eq{freeenergy} is peculiar in the 
sense that it is duality invariant purely with the term proportional to
the volume. This fact may imply that the local degrees of freedom of the 
hypothetical system is given purely by the proposal of Atick-Witten without
any complications of non-locality.
Assuming this volume-dilaton dependence, an expression of the action
with the reparameterization invariance of $t$ must
be in the form 
\be
\label{formh}
\int dt\ \sqrt{-g_{00}}\ e^{-\varphi}h(T/\sqrt{-g_{00}}),
\ee
where $h$ is an arbitrary function. $\varphi$ is not invariant under the 
S-duality \eq{sduality}, and to cancel the variation by a $g_{00}$ dependence, 
the unique choice is the form $h(x)\sim x^2$. Thus the expression 
\eq{freeenergy} can be obtained from the assumption that the free energy be 
proportional to the volume and invariant under 
the T- and S-dualities.

It is also interesting to look at \eq{freeenergy} from the M-theory viewpoint
\cite{Townsend:1995kk,Witten:1995ex}.
The type IIA string theory is obtained by compactifying one of the M-theory 
spatial dimensions, say the eleventh dimension. The relation between the 
M-theory metric and the string metric is given by \cite{Townsend:1996xj}
\bea
ds_M^2&=&g^M_{\mu\nu}dx^\mu dx^\nu
=e^{-\frac23 \phi} ds_{string}^2+e^{\frac43\phi} dy^2, 
\eea
where $y$ is the eleventh dimension.
Rewriting with the M-theory metric, the action \eq{hypoaction} becomes
\be
S_M=-\int d^{11}x\ \sqrt{-g^M} \left(\frac{T}{\sqrt{-g^M_{00}}}\right)^2.
\ee
Therefore, the free energy of the hypothetical system in M-theory is 
\be
\label{freem}
F_M=-V_{10} T^2,
\ee
where $V_{10}$ is the ten-dimensional spatial volume of M-theory.
This expression shows that the eleventh dimension is treated
equivalently with the other spatial dimensions. Hence the expression
\eq{freeenergy} is invariant under the exchange of the eleventh dimension
with the compactified dimensions of string theory.  
It is also clear that, if I impose this exchange symmetry on \eq{formh},
the unique choice of the free energy is \eq{freeenergy}. 
This implies that I may impose this exchange symmetry instead of S-duality
to obtain the expression \eq{freeenergy} from the assumption that the free 
energy be proportional to volume. 
Amusingly the expression \eq{freem} suggests that, the local degrees of 
freedom of M-theory behaves like a (1+1)-dimensional field theory as  
string theory rather than (1+2) as would be a natural expectation from that 
M-theory is often referred as membrane theory.   
Since all the spatial directions are totally equivalent in \eq{freem},  
it would be hard to imagine that the temperature dependence has been modified 
in the process of compactifying the eleventh dimension to obtain 
the IIA string theory.

In this paper, I have shown how unique the expressions \eq{entropybound} 
and \eq{freeenergy} are in the viewpoint of string/M-theory.
Comparing with the macroscopic derivations of these formulas, 
it seems surprising for these formulas 
to satisfy the microscopic requirements of dualities.
But are there any chances to obtain the expression \eq{freeenergy} directly 
in string theory? As explicitly shown in \cite{Atick:1988si}, 
the Hagedorn transition is caused by
the instability of a stringy mode which winds around the compact temperature 
direction. The mode will condense above the Hagedorn transition and
allow genus-zero contributions to appear by making tiny holes with 
non-zero winding numbers on string world sheets. But in a low order 
approximation of string theory effective action, there were no stable minima 
and it was impossible to evaluate the free energy above the Hagedorn 
transition \cite{Atick:1988si}. 
On the other hand, assuming this story and the existence of 
a stable minimum, because the condensation is spatially local, 
the free energy would be dominated by a genus-zero contribution proportional 
to the volume, 
\be
F\sim e^{-2\phi}V h(T),
\ee
where $h(T)$ is a function of the temperature. 
This takes the same form as \eq{formh}, and, as argued above,  
if I impose the S-duality \eq{sduality} or the exchange symmetry of 
the eleventh dimension of M-theory,
the free energy is constrained to the form \eq{freeenergy}.
Thus there seems to exist a good chance of obtaining the free energy 
expression 
\eq{freeenergy} by a string theory computation which respects the duality
symmetries and is not limited by the string world sheet picture.
This high requirement of a non-perturbative formulation of string theory
seems to make the formula \eq{freeenergy} a fascinating primary goal for 
understanding string theory beyond the Hagedorn transition. 
But this line of thought is obscured by the 
fact that, if the relevant temperature is 
the order of or above the Hagedorn transition,  
there will be no controllable parameters to slow down the background 
evolution caused by the genus-zero contributions and the static approximation 
justifying the thermodynamic treatment will be no longer valid 
\cite{Atick:1988si}.



\begin{thebibliography}{99}
\bibitem{Atick:1988si}
J.~J.~Atick and E.~Witten,
``The Hagedorn Transition And The Number 
Of Degrees Of Freedom Of String Theory,''
Nucl.\ Phys.\ B {\bf 310}, 291 (1988).

\bibitem{Polchinski:rq}
See also Chapter 9 of J.~Polchinski,
``String Theory. Vol. 1: An Introduction To The Bosonic String,''
{\it  Cambridge, UK: Univ. Pr. (1998) 402 p}.

\bibitem{Bekenstein:jp}
J.~D.~Bekenstein,
``A Universal Upper Bound On The Entropy To Energy Ratio For Bounded Systems,''
Phys.\ Rev.\ D {\bf 23}, 287 (1981).

\bibitem{'tHooft:gx}
G.~'t Hooft,
``Dimensional Reduction In Quantum Gravity,''
arXiv:gr-qc/9310026.

\bibitem{Susskind:1994vu}
L.~Susskind,
``The World as a hologram,''
J.\ Math.\ Phys.\  {\bf 36}, 6377 (1995)
[arXiv:hep-th/9409089].

\bibitem{Flanagan:1999jp}
E.~E.~Flanagan, D.~Marolf and R.~M.~Wald,
``Proof of Classical Versions of the Bousso Entropy Bound 
and of the Generalized Second Law,''
Phys.\ Rev.\ D {\bf 62}, 084035 (2000)
[arXiv:hep-th/9908070].

\bibitem{Sasakura:1999xp}
N.~Sasakura,
``An uncertainty relation of space-time,''
Prog.\ Theor.\ Phys.\  {\bf 102}, 169 (1999)
[arXiv:hep-th/9903146].

\bibitem{Fischler:1998st}
W.~Fischler and L.~Susskind,
``Holography and cosmology,''
arXiv:hep-th/9806039.

\bibitem{Bousso:1999xy}
R.~Bousso,
``A Covariant Entropy Conjecture,''
JHEP {\bf 9907}, 004 (1999)
[arXiv:hep-th/9905177];
``Holography in general space-times,''
JHEP {\bf 9906}, 028 (1999)
[arXiv:hep-th/9906022].

\bibitem{Veneziano:1999ty}
G.~Veneziano,
``Pre-bangian origin of our entropy and time arrow,''
Phys.\ Lett.\ B {\bf 454}, 22 (1999)
[arXiv:hep-th/9902126];
R.~Brustein and G.~Veneziano,
``A Causal Entropy Bound,''
arXiv:hep-th/9912055.

\bibitem{Verlinde:2000wg}
E.~Verlinde,
``On the holographic principle in a radiation dominated universe,''
arXiv:hep-th/0008140.

\bibitem{Banks:2001px}
T.~Banks and W.~Fischler,
``An holographic cosmology,''
arXiv:hep-th/0111142.

\bibitem{T-duality}
K.~Kikkawa and M.~Yamasaki,
 ``Casimir Effects In Superstring Theories,''
Phys.\ Lett.\ B {\bf 149}, 357 (1984);
N.~Sakai and I.~Senda,
``Vacuum Energies Of String Compactified On Torus,''
Prog.\ Theor.\ Phys.\  {\bf 75}, 692 (1986)
[Erratum-ibid.\  {\bf 77}, 773 (1987)].

\bibitem{Polchinski}
See for example, J.~Polchinski,
``String Theory. Vol. 1: An Introduction To The Bosonic String,''
{\it  Cambridge, UK: Univ. Pr. (1998) 402 p};
``String Theory. Vol. 2: Superstring Theory And Beyond,''
{\it  Cambridge, UK: Univ. Pr. (1998) 531 p}.

\bibitem{Tseytlin:1991xk}
A.~A.~Tseytlin and C.~Vafa,
``Elements of string cosmology,''
Nucl.\ Phys.\ B {\bf 372}, 443 (1992)
[arXiv:hep-th/9109048].

\bibitem{Townsend:1995kk}
P.~K.~Townsend,
``The eleven-dimensional supermembrane revisited,''
B {\bf 350}, 184 (1995)
[arXiv:hep-th/9501068].

\bibitem{Witten:1995ex}
E.~Witten,
``String theory dynamics in various dimensions,''
arXiv:hep-th/9503124.

\bibitem{Townsend:1996xj}
P.~K.~Townsend,
``Four lectures on M-theory,''
in {\it C96-06-10.2}
arXiv:hep-th/9612121.

\end{thebibliography}
\end{document}